\documentstyle[12pt,epsf]{article}

\begin{document}
\baselineskip 0.7cm

\newcommand{\beq}{ \begin{eqnarray} }
\newcommand{\eeq}{ \end{eqnarray} }
\newcommand{\beqstar}{ \begin{eqnarray*} }
\newcommand{\eeqstar}{ \end{eqnarray*} }
\newcommand{\gsim}{ \mathop{}_{\textstyle \sim}^{\textstyle >} }
\newcommand{\lsim}{ \mathop{}_{\textstyle \sim}^{\textstyle <} }
\newcommand{\vev}[1]{ \left\langle {#1} \right\rangle }
\newcommand{\lsp}{ \left ( }
\newcommand{\rsp}{ \right ) }
\newcommand{\lmp}{ \left \{ }
\newcommand{\rmp}{ \right \} }
\newcommand{\llp}{ \left [ }
\newcommand{\rlp}{ \right ] }
\newcommand{\labs}{ \left | }
\newcommand{\rabs}{ \right | }
\newcommand{\EV}{ {\rm eV} }
\newcommand{\KEV}{ {\rm keV} }
\newcommand{\MEV}{ {\rm MeV} }
\newcommand{\GEV}{ {\rm GeV} }
\newcommand{\TEV}{ {\rm TeV} }
\newcommand{\mgut}{M_{\rm GUT}}
\newcommand{\mint}{M_{I}}
\newcommand{\mll}{m_{\tilde{l}L}^{2}}
\newcommand{\mdr}{m_{\tilde{d}R}^{2}}
\newcommand{\mllXX}[1]{m_{\tilde{l}L , {#1}}^{2}}
\newcommand{\mdrXX}[1]{m_{\tilde{d}R , {#1}}^{2}}
\newcommand{\mgy}{m_{G1}}
\newcommand{\mgl}{m_{G2}}
\newcommand{\mgc}{m_{G3}}
\newcommand{\nuR}{\nu_{R}}
\newcommand{\slL}{\tilde{l}_{L}}
\newcommand{\slLi}{\tilde{l}_{Li}}
\newcommand{\sdR}{\tilde{d}_{R}}
\newcommand{\sdRi}{\tilde{d}_{Ri}}

\renewcommand{\thefootnote}{\fnsymbol{footnote}}
\setcounter{footnote}{1}

\begin{titlepage}

\begin{flushright}
TU-476
\\
January, 1995
\end{flushright}

\vskip 0.35cm
\begin{center}
{\large \bf
Lepton-Flavor Violation in the Supersymmetric Standard Model
with Seesaw-Induced Neutrino Masses
}
\vskip 1.2cm
J.~Hisano, T.~Moroi, K.~Tobe, M.~Yamaguchi and T.~Yanagida
\vskip 0.4cm

{\it Department of Physics, Tohoku University,\\
     Sendai 980-77, Japan}

\vskip 1.5cm

\abstract{
We examine the lepton-flavor violation caused by a Yukawa coupling
matrix $y_{\nu,ij}$ for right-handed neutrinos in the supersymmetric
standard model. We stress that decay rates for
$\tau\rightarrow\mu\gamma$ and $\mu\rightarrow e\gamma$ may reach the range
to be accessible to near future experiments if left-right mixing terms
in the slepton mass matrix are substantially large.
}
\end{center}
\end{titlepage}

\renewcommand{\thefootnote}{\arabic{footnote}}
\setcounter{footnote}{0}

%
%
%
%

A small, but non-vanishing neutrino mass, if any, is regarded as an
important indication of a new physics beyond the standard model. The
most interesting candidate is the well-known seesaw
model~\cite{seesaw} that explains very naturally the small mass
for neutrino in terms of a large Majorana mass for a right-handed
neutrino $N$. This model has attracted many authors not only because
of its natural structure but also because the presence of $N$ would
illuminate some of the deep questions in particle physics.

{}From the phenomenological point of view, on the other hand, introduction
of three families of the right-handed neutrinos $N^i$ (where $i$ and $j$
are flavor indices) brings two new ingredients to the standard model;
one is a new scale of the Majorana masses for $N^i$, $\mint$, and the
other a new matrix for Yukawa coupling constants of $N^i$, $y_{\nu
,ij}$.  Thus, we have two independent Yukawa matrices in the lepton
sector as in the quark sector.  In general, a simultaneous
diagonalization of the both matrices is quite accidental and then the
addition of the new Yukawa coupling $y_{\nu ,ij}$ causes a lepton-flavor
violation.  In the standard model, however, the amplitudes for the
lepton-flavor violating processes at low energies are suppressed by an
inverse power of $\mint$ at least and hence we do not expect sizable
rates for such processes as far as $\mint$ is very large. The neutrino
oscillation is a famous exception in this point of view.

It has been, however, noted by several authors~\cite{PRL57-961} that in
the supersymmetric standard model (SSM) the Yukawa coupling $y_{\nu
,ij}$ of $N^i$ generates off-diagonal entries in the mass matrices for
sleptons through the renormalization effects, which leads to
unsuppressed lepton-flavor violations such as
$\tau\rightarrow\mu\gamma$, $\mu\rightarrow e\gamma$, and so
on.\footnote
{It has been pointed out very clearly~\cite{PLB338-212} that the similar
lepton-flavor violation occurs in the SUSY grand unified theories
(GUTs).}
The predicted rates for the individual processes depend on the unknown
Yukawa matrix $y_{\nu ,ij}$.  If off-diagonal elements of $y_{\nu ,ij}$
are very small like in the quark sector, however, the reaction rates for
these processes are predicted too small to be accessible to the near
future experiments.

The purpose of this letter is to point out that left-right mixing terms
in the slepton mass matrix may give larger contributions to the
lepton-flavor violating processes, which have not been considered in the
previous literatures~\cite{PRL57-961}. We show, as a consequence, that
the decay rates for $\tau\rightarrow\mu\gamma$ and $\mu\rightarrow
e\gamma$ can reach indeed the range close to the present experimental
upper limits ${\rm Br}(\tau\rightarrow\mu\gamma)\leq 4.2\times 10^{-6}$
and ${\rm Br}(\mu\rightarrow e\gamma)\leq 4.9\times 10^{-11}$~\cite{PDG}
even when the off-diagonal elements of $y_{\nu ,ij}$ are small as
$y_{\nu ,23}/y_{\nu ,33}\sim 0.04$ and $y_{\nu ,13}/y_{\nu ,33}\sim
0.01$.

In the SSM with three families of right-handed neutrinos $N^i$ ($i$ =
1 -- 3), the superpotential $W$ for the lepton sector is given by,
\beq
W = y_{e, ij} \bar{E}^i L^j H_1 + y_{\nu ,ij} N^i L^j H_2
+\frac{1}{2} M_{I,ij} N^iN^j,
\label{superpot}
\eeq
where $L^i$, $\bar{E}^i$ and $N^i$ are the chiral multiplets for
left-handed lepton doublets, right-handed lepton singlets and
right-handed neutrinos, and $H_1$ and $H_2$ those for the Higgs
doublets. For simplicity, we assume the mass matrix $M_{I,ij}$ is
proportional to the unit matrix as
\beq
M_{I,ij} = \mint
\lsp \begin{array}{ccc} 1 & & \\ & 1 & \\ & & 1 \end{array} \rsp
\label{majorana_mass}
\eeq
The analysis with the general case $M_{I,ij} = M_{I,i}\delta_{ij}$ is
straightforward, but as long as $M_{Ii}\sim M_{Ij} $ the conclusion will
not be much different from that in the simplest case (\ref{majorana_mass}).
We choose a basis on which $y_{e,ij}$ is diagonal at the electroweak  scale
$m_Z$,
\beq
y_{e,ij}(\mu = m_Z) =  y_{e,i} \delta_{ij}.
\eeq

To demonstrate our main point, we assume tentatively that the Yukawa
couplings of $N$ are identical to those of the up-type
quarks, similar to the case of SO(10) unification,\footnote{We
do not use the other SO(10)-like relation $y_{e,ij} =y_{d,ij}$ for
the down-type quarks, since it gives wrong results on the masses for the
first and second families.  }
\begin{equation}
 y_{\nu ,ij} = y_{u,ij}.
\label{yukawa-up}
\end{equation}
 We suppose that these relations hold at the gravitational scale $\mu
=M_G\simeq 2.4\times 10^{18}\GEV$.
With the basis where the Yukawa couplings for the down-type quarks are
flavor-diagonal at the electroweak scale, $y_{d,ij}=y_{d,i}
\delta_{ij}$,  the $y_{u,ij}$ is written as
\begin{equation}
y_{u,ij}=  V_{\rm KM}^T
\lsp \begin{array}{ccc}
y_u & & \\ & y_c & \\ & & y_t
\end{array} \rsp
V_{\rm KM},
\end{equation}
where $V_{\rm KM}$ is the Kobayashi-Maskawa matrix.  The meaning of
the Yukawa couplings of the up-type quarks $y_u$, $y_c$ and $y_t$ is
the obvious one.

The seesaw mechanism~\cite{seesaw} induces small neutrino masses
as\footnote{This equation holds only at tree-level.  In our numerial
analysis, we take into account the renormalization effects.}
\beq
m_{\nu ,ij} &\simeq&
y^T_{\nu ,ik}\frac{1}{M_{I,k}}y_{\nu ,kj} \vev{H_2}^2
\nonumber \\
&=& \frac{\vev{H_2}^2}{\mint} V_{\rm KM}^T
\lsp \begin{array}{ccc} y_u& & \\ & y_c & \\ & & y_t \end{array} \rsp
V_{\rm KM} V_{\rm KM}^T
\lsp \begin{array}{ccc} y_u& & \\ & y_c & \\ & & y_t \end{array} \rsp
V_{\rm KM}.
\eeq
We see  that the mixing matrix appearing in the neutrino
oscillation is approximately identical with $V_{\rm KM}$, due to the
large hierarchy $y_u\ll y_c\ll y_t$~\cite{PLB97-99}. For a given
$\mint$, we determine the mass eigenvalues and the mixing angles for the
neutrinos.  We adjust $M_I\sim 10^{12}\GEV$ so that the  mass
for $\nu_\tau$ is predicted to be  $m_{\nu_\tau} =
10\EV$,\footnote
{In this case, the mass for $\nu_\mu$ is predicted as $\sim
10^{-3}\EV$, which is the right value for the MSW solution to the
solar neutrino problem~\cite{MSW}. However, the predicted mixing angle
$\theta_{\nu_e\nu_\mu}\simeq\theta_{\rm Cabibbo}$ is much larger than
that in the small angle solution of MSW~\cite{Fukugita&Yanagida}.
Therefore, to account for the solar neutrino deficit, we must use
smaller value for $y_{\nu ,12}$.  However, our main conclusions for
$\tau\rightarrow\mu\gamma$ and $\mu\rightarrow e\gamma$ given in the
text are unchanged, since the Yukawa coupling $y_{\nu 12}$ is almost
irrelevant for these processes as noted in ref.~\cite{PLB338-212}.}
which lies in an interesting range in
cosmology~\cite{Kolb&Turner}.
To estimate the Yukawa matrix $y_{u,ij}(\mu =M_G)$ we use the observed
values $m_u(1\GEV)=4.5\MEV$, $m_c(1\GEV)=1.27\GEV$~\cite{PRepC87-77},
$m_t=174\GEV$~\cite{PRD50-2966} and $V_{\rm KM}$.  In our numerical
calculations, we use the central value of each matrix element
$V_{{\rm KM},ij}$ given in ref.\cite{PDG}.

We are now in a position to discuss the lepton-flavor violation. In the
minimal SUSY standard model (MSSM), soft SUSY-breaking mass terms for
sleptons have the general form,
\beq
{\cal L}_{\rm soft} =
- m_{E, ij}^2 \widetilde{\overline{E}}^i \widetilde{\overline{E}}^{j\dagger} -
m_{L, ij}^2 \widetilde{L}^{i\dagger} \widetilde{L}^j
- \lsp A_{e,ij} \widetilde{\overline{E}}^i \widetilde{L}^j H_1 + h.c. \rsp.
\eeq
The lepton-flavor conservation is easily violated by taking
non-vanishing off-diagonal elements of each matrices and the sizes of
such elements are strongly constrained from experiments. In the SUSY
standard model based on the supergravity~\cite{NPB212-413}, it is
therefore assumed that the mass matrices $m_E^2$ and $m_L^2$ are
proportional to the unit matrix, and $A_{e,ij}$ is proportional to the
Yukawa matrix $y_{e,ij}$. With these soft terms, the lepton-flavor
number is conserved exactly.

It is, however, not true if the effects of the right-handed neutrinos
are taken into account~\cite{PRL57-961}. The Yukawa coupling of neutrino
$y_{\nu ,ij}N^iL^jH_2$ in eq.(\ref{superpot}) and the soft SUSY-breaking
terms such as
\beq
{\cal L}_{\rm soft,\nu} &=&
- m_{N, ij}^2 \widetilde{N}^i \widetilde{N}^{j\dagger}
- \lsp A_{\nu ,ij} \widetilde{N}^i \widetilde{L}^j H_2 + h.c. \rsp
\eeq
induce off-diagonal elements of $m_L^2$ through the radiative
corrections. The renormalization group equation (RGE) for $m_L^2$ is
given by
\beq
\frac{dm_{L,ij}^2}{d\ln\mu}
&=& \lsp\frac{dm_{L,ij}^2}{d\ln\mu}\rsp_{\rm MSSM} +
\frac{1}{16\pi^2} \Big\{ (m_L^2 y_\nu^\dagger y_\nu)_{ij} +
(y_\nu^\dagger y_\nu m_L^2)_{ij} + 2(y_\nu^\dagger y_\nu)_{ij} m_{H2}^2
\nonumber \\ &&
+ (y_\nu^\dagger m_N^2 y_\nu)_{ij} + 2(A_\nu^\dagger A_\nu)_{ij}\Big\},
\label{rge4ml2}
\eeq
where $(dm_{L,ij}^2/d\ln\mu)_{\rm MSSM}$ represents the
$\beta$-function in  the MSSM,\footnote
{See ref.~\cite{PRD50-2282} for explicit formulae.}
and $m_{H2}^2$ is the soft SUSY breaking mass for the Higgs doublet
$H_2$. As one can see easily in eq.(\ref{rge4ml2}), off-diagonal
elements of $m_{L,ij}^2$ are induced by the renormalization effects if
non-vanishing off-diagonal elements of $y_\nu$ and $A_\nu$ exist.

In order to obtain the slepton mass matrix $m_{L,ij}^2$ at the
electroweak scale, we solve the RGEs for the full relevant parameters
numerically.  At the energy scale $\mint\leq\mu\leq M_G$ we use the
RGEs derived from the MSSM with the right-handed neutrinos, and below
the energy scale $\mint$ the MSSM-RGEs without the right-handed
neutrinos. For the soft SUSY-breaking parameters, we assume the
following boundary conditions suggested in the minimal
supergravity~\cite{NPB212-413,PRep110-1};
\beq
m_{f,ij}^2 (\mu =M_G) &=& m_0^2 \delta_{ij},
\label{universalmass}
\\
A_{f,ij} (\mu =M_G) &=& a~m_0~y_{f,ij},
\label{BC4A}
\eeq
where $m_{f,ij}^2$ is the soft SUSY-breaking mass matrix for sfermion
$\tilde{f}$ (with $f=u$, $d$, $e$ and $\nu$), $A_{f,ij}$ the so-called
$A$-parameter, $y_{f,ij}$ the Yukawa coupling constants for the fermion
$f$, $m_0$ the universal SUSY-breaking mass, and $a$ is a free
parameter of $O(1)$. For gaugino masses, we use tentatively the
relation implied by the SUSY GUTs \cite{PRep110-1}
\beq
\frac{m_{G3}}{g_3^2} = \frac{m_{G2}}{g_2^2}
= \frac{3}{5}\times\frac{m_{G1}}{g_1^2} ,
\label{GUT-relation}
\eeq
where $m_{G3}$, $m_{G2}$, $m_{G1}$, and $g_3$, $g_2$, $g_1$ are the
gaugino masses and gauge coupling constants for the gauge groups
SU(3)$_C$, SU(2)$_L$ and U(1)$_Y$, respectively.

The soft SUSY breaking mass matrix $m_{L,ij}^2$ is determined by
solving the coupled RGEs for all relevant parameters. Since
$\tan\beta(\equiv\langle H_2\rangle/\langle H_1\rangle)$ dependence of
the result is relatively mild, we take $\tan\beta =3$ for the time
being. The result for the larger $\tan\beta$ will be given later for
comparison.  For the case $0\leq m_{G2}\leq m_0$, the mass matrix
$m_{L,ij}^2$ at the electroweak scale is given by
\beq
m_{L}^2(\mu=m_Z) &\simeq&
\lsp \begin{array}{ccc}
(1.0 - 2.0) & -(0.9 - 1.1)\times 10^{-4} & -(2.3-2.7)\times 10^{-3} \\
-(0.9-1.1)\times 10^{-4} & (1.0 - 2.0) & -(1.0 - 1.2)\times 10^{-2} \\
-(2.3-2.7)\times 10^{-3} & -(1.0 - 1.2)\times 10^{-2} & (0.75 - 1.69)
\end{array} \rsp
\nonumber \\ &&
\times m_0^2,
\label{m_L^2@mz}
\eeq
where the magnitude of each entry becomes larger as $m_{G2}$ larger.
However, for the case $m_{G2}(\mu =m_Z)> m_0$, ratios of the
off-diagonal elements to the diagonal ones become smaller than those in
eq.(\ref{m_L^2@mz}), which give a more suppression of the lepton-flavor
violation as a result.

We are now ready to calculate the reaction rates for various
lepton-flavor violating processes. Let us first consider the diagrams
considered in ref.\cite{PRL57-961} contributing to the process
$l_i\rightarrow l_j\gamma$.  In this paper, we ignore the mixings in
the neutralino and chargino sectors and consider that the bino and the
winos are mass eigenstates.  This approximation is justified if the
SUSY-invariant higgsino mass $\mu_H$ is large, which is nothing but a
situation we will consider in the present paper.

{}From the diagrams in fig.~\ref{fig:feyn_LL}, we obtain the amplitude
for this process as
\beq
{\cal M} (l_i\rightarrow l_j\gamma) =
C_{LL} \overline{l}_j P_R [\not{q},\not{\epsilon}] l_i,
\label{M_LL}
\eeq
where $q$ and $\epsilon$ are the momentum and polarization vector
of the emitted photon. Here, the coefficient $C_{LL}$ is given by
\beq
C_{LL} &=&
\frac{1}{16\pi^2} e m_{l_i} \Bigg\{
\frac{m_{L,ij}^2}{m_{\tilde{e}_L}^4}
g_1^2 S_e (m_{G1}^2/m_{\tilde{e}_L}^2)
\nonumber \\ &&
+ \frac{m_{L,ij}^2}{m_{\tilde{e}_L}^4}
g_2^2 S_e (m_{G2}^2/m_{\tilde{e}_L}^2)
+ \frac{m_{L,ij}^2}{m_{\tilde{\nu}_L}^4}
g_2^2 S_\nu (m_{G2}^2/m_{\tilde{\nu}_L}^2)
\Bigg\},
\eeq
where $m_{l_i}$, $m_{\tilde{e}_L}$ and $m_{\tilde{\nu}_L}$ represent
masses for the lepton $l_i$, charged $\tilde e_L$ and neutral sleptons
$\tilde \nu_L$,
respectively,\footnote
{We have assumed $m_{l_i}\gg m_{l_j}$.}
and the functions $S_e$ and $S_\nu$ are given by
\beq
S_e(x) &=& \frac{1}{48(x-1)^5}
\{ 17x^3 -9x^2 -9x +1 -6x^2(x+3)\ln x \},
\\
S_\nu(x) &=& -\frac{1}{12(x-1)^5}
\{ -x^3 -9x^2 +9x +1 +6x(x+1)\ln x \}.
\eeq
Then, the decay rate can be obtained from the amplitude (\ref{M_LL}) as
\beq
\Gamma (l_i\rightarrow l_j\gamma) =
\frac{1}{4\pi} \labs C_{LL} \rabs^2 m_{l_i}^3.
\label{rate-ll}
\eeq
By using the $m_{L,ij}^2$ given in eq.(\ref{m_L^2@mz}), we have
calculated the branching ratios of the processes
$\tau\rightarrow\mu\gamma$ and $\mu\rightarrow e\gamma$. We have
checked that for $m_{G2}=45\GEV$, the branching ratios for each
processes have the maximums $O(10^{-9})$ and $O(10^{-12})$,
respectively, at the possible minimum value of the slepton mass $\simeq
45\GEV$, which are, however, much smaller than the present
experimental limits (${\rm Br}(\tau\rightarrow\mu\gamma)\leq 4.2\times
10^{-6}$ and ${\rm Br}(\mu\rightarrow e\gamma)\leq 4.9\times
10^{-11}$~\cite{PDG}).

So far, we have neglected the left-right mixings in the slepton mass
matrix. However, if the SUSY invariant mass $\mu_H$ for Higgs doublets
is so big that  $m_\tau\mu_H \tan\beta\sim O(m_0^2)$, we have non-negligible
left-right mixings in the slepton mass matrix. As we will show below,
these mixing terms give rise to contributions to the $l_i\rightarrow
l_j\gamma$ process larger than the previous estimate in
eq.(\ref{rate-ll}). The main reason for this is that the chirality flip
in the fermion line occurs at the internal gaugino mass term (see
fig.~\ref{fig:feyn_LR}) while in the previous case it occurs at the
external lepton mass term. Since the gaugino mass is much bigger than
the lepton mass, these new diagrams may yield dominant contributions.

Before calculating the diagrams in fig.~\ref{fig:feyn_LR}, we first
diagonalize the charged slepton mass matrix given by the following
$6\times 6$ matrix;
\beq
{\cal L}_{\rm mass} = - (\widetilde{L}^\dagger , \widetilde{\bar{E}})
\lsp \begin{array}{cc}
m_{LL}^2 & m_{LR}^2 \\
m_{LR}^{2\dagger} & m_{RR}
\end{array} \rsp
\lsp \begin{array}{c}
\widetilde{L} \\ \widetilde{\bar{E}}^\dagger
\end{array} \rsp,
\label{mass-matrix}
\eeq
with
\beq
m_{LL,ij}^2 &=&
m_{L,ij}^2
+ m_Z^2\cos 2\beta\lsp\sin^2\theta_W -\frac{1}{2}\rsp\delta_{ij},
\label{mass-matrix-ll}
\\
m_{RR,ij}^2 &=& m_{E,ij}^2 - m_Z^2\cos 2\beta \sin^2\theta_W
\delta_{ij},
\label{mass-matrix-rr}
\\
m_{LR,ij}^2 &=& m_{l_i} \delta_{ij} \mu_H \tan\beta + A_{E,ji} v
\cos\beta/\sqrt{2}
\eeq
where $m_Z$ is the $Z$-boson mass and $\sin^2\theta_W$ the Weinberg
angle. Notice that as far as the $a$-parameter in eq.(\ref{BC4A}) is $O(1)$,
the $A_{E,ij}$ is negligibly small in  the case of  $\mu_H\tan\beta\gg
m_0$  which concerns us. $m_{E,ij}^2$ at the electroweak scale are
determined by solving the RGEs in the same way of the previous
calculation for $m_{L,ij}^2$, and are found to be $m_{E,ij}^2 \simeq
(1.0-1.4)m_0^2\delta_{ij}$.

With the diagonalization matrix $U$ for eq. (\ref{mass-matrix}) and the
eigenvalues $m_{\tilde{e},A}^2$ ($A$ = 1 -- 6), we can write the
amplitude for the process $l_i\rightarrow l_j\gamma$ as
\beq
{\cal M} (l_i\rightarrow l_j\gamma) =
C_{LR} \overline{l}_j P_R [\not{q},\not{\epsilon}] l_i,
\label{M_LR}
\eeq
where the coefficient $C_{LR}$ is given by
\beq
C_{LR} = \frac{1}{16\pi^2} e g_1^2 m_{G1}
\sum_A U_{jA} U_{A i+3}^\dagger \frac{1}{m_{\tilde{e},A}^2}
T_{LR}(m_{G1}^2/m_{\tilde{e},A}^2),
\label{ampLR}
\eeq
with
\beq
T_{LR}(x) = \frac{1}{4(x-1)^3} (x^2 - 1 - 2x\ln x).
\eeq
We find that in the case of small $m_{LR}^2$, this amplitude
(\ref{ampLR}) is well approximated by
\begin{eqnarray}
\left. C_{LR} \right|_{\rm mass-insertion} &= &
  \frac{1}{16 \pi^2}e g_1^2 m_{G1} \sum_k U_{Ljk} U^\dagger_{Lki}
   (m_{l_i} \mu_H \tan\beta+ A_{E,ii} v \cos\beta/\sqrt{2})
\nonumber \\
 &   \times & \frac{1}{m_{\tilde e_{Ri}}^2-m_{\tilde e_{Lk}}^2}
           \left [ \frac{1}{m_{\tilde e_{Ri}}^2}
                   T_{LR} \left( \frac{m_{G1}^2}{m_{\tilde e_{Ri}}^2}
                   \right)
              -     \frac{1}{m_{\tilde e_{Lk}}^2}
                   T_{LR}  \left ( \frac{m_{G1}^2}{m_{\tilde e_{Lk}}^2}
                   \right ) \right ],
\label{ampLR-mass-insertion}
\end{eqnarray}
which is obtained by using the $m_{LR}^2$ mass insertion.  Here,
$U_L$ is the diagonalization matrix for the mass matrix of the
left-handed sleptons (\ref{mass-matrix-ll}), $m_{\tilde e_{Lk}}^2$ its
eigenvalues and $m_{\tilde e_{Ri}}^2 = m_{RR,ii}^2$ in
eq.(\ref{mass-matrix-rr}). From the amplitude (\ref{M_LR}), we get the
decay rate as
\beq
\Gamma (l_i\rightarrow l_j\gamma) =
\frac{1}{4\pi} \labs C_{LR} \rabs^2 m_{l_i}^3,
\label{rate-lr}
\eeq
which should be compared with the result in eq.(\ref{rate-ll}).

For a given set of $\tan\beta$ and $m_{G2}$, the decay rates depend
basically on the parameters $\mu_H$ and $m_0$. However, it is convenient
to express them in terms of two physical parameters, $m_{\tilde
\tau_1}$ and $m_{\tilde \tau_2}$ ($m_{\tilde \tau_1} < m_{\tilde
\tau_2}$), which are mass eigenvalues for the sleptons belonging
mainly to the $\tau$-family.  The results are shown in
fig.~\ref{fig:br_light} for $\tan\beta =3$, with $m_{G2}=45\GEV$ and
$m_{\tilde{\tau}_1}=50\GEV$ being fixed. For a comparison, we also show
the results for the case of $\tan\beta =30$. We easily see that the
branching ratios for $\tau\rightarrow\mu\gamma$ and $\mu\rightarrow
e\gamma$ are predicted as
\beq
{\rm Br}(\tau\rightarrow\mu\gamma) &\simeq&
3\times 10^{-8} - 4\times 10^{-7},
\label{BR_tau} \\
{\rm Br}(\mu\rightarrow e\gamma) &\simeq&
5\times 10^{-12} - 2\times 10^{-11},
\label{BR_mu}
\eeq
for $m_{\tilde{\tau}_2} \simeq (100-250)\GEV$.\footnote
{The minimum value ($m_{\tilde \tau_2} \simeq 100 \GEV$) comes from
the constraint that the lightest sneutrino should be heavier than
$41.7\GEV$~\cite{PDG}.}
It should be stressed that the large branching ratios in
eqs.(\ref{BR_tau}) and (\ref{BR_mu}) are never obtained in the case
where the left-right mass-insertion is applicable
($\mu_H\tan\beta\lsim m_0$). To see how the results depend on
$m_{\tilde{\tau}_1}$ and the gaugino mass, we show our results in
fig.~\ref{fig:br_heavy} taking $m_{\tilde{\tau}_1}=100\GEV$ and
$m_{G2}=90\GEV$. From figs.~\ref{fig:br_light} and
\ref{fig:br_heavy}, we see that the obtained branching ratios are much
larger than the previous estimates from eq.(\ref{rate-ll}) and
they lie in the range which will be studied experimentally in no
distant future.

We should note that our results shown in figs.~\ref{fig:br_light} and
\ref{fig:br_heavy} are also larger than the SUSY-GUT
predictions~\cite{PLB338-212}. This is because the authors in
ref.~\cite{PLB338-212} have not taken into account the left-right mixing
effects. However, if these effects are induced, the similar conclusion
may be obtained even in the SUSY SU(5) GUTs.

We now briefly discuss other lepton-flavor violating processes such as
$\mu\rightarrow 3e$ and  $\tau\rightarrow 3\mu$.  With the
parameter space discussed in the present paper, there also
appears an enhancement factor $m_{G2}/m_{\tau}$ in the Penguin
diagrams, which will give dominant contributions to these processes.
In this case, the decay rates of $\mu \rightarrow 3 e$ and $\tau
\rightarrow 3 \mu$ have simple relations to those of the
$\mu\rightarrow e\gamma$ and $\tau \rightarrow \mu \gamma$ processes as
\begin{eqnarray}
\frac{{\rm Br}(\mu\rightarrow 3e)|_{\rm Penguin}}
{{\rm Br}(\mu\rightarrow e\gamma)}& \simeq & \frac{2 \alpha}{3 \pi}
 \ln\frac{m_{\mu}}{ m_e}  \simeq 0.8 \times
10^{-2},
\\
\frac{{\rm Br}(\tau \rightarrow 3 \mu)|_{\rm Penguin}}
{{\rm Br}(\tau \rightarrow \mu \gamma)}& \simeq & \frac{2 \alpha}{3 \pi}
\ln\frac{m_{\tau}}{ m_{\mu}} \simeq 0.4 \times
10^{-2},
\end{eqnarray}
where we have taken only the logarithmic contributions.
As for the $\mu$-e conversion in nuclei, the amplitudes of the box
diagrams depend  on the squark masses. Thus, in the
large-squark-mass region, the box diagrams yields negligible
contributions. The detailed analysis on various lepton-flavor violating
processes including the $\mu$-e conversion will be given in the future
publication~\cite{future}.

Several comments are in order.

\noindent
{\it i}) We have considered the
parameter space where $m_{\tilde{\tau}_1}\simeq (50-100)\GEV$
$m_{\tilde{\tau}_2}\simeq (100-250)\GEV$ and $\tan\beta =3-30$. These
parameters correspond to the original parameters $\mu_H$ and $m_0$
as
\beq
\mu_H =
\lmp \begin{array}{ll}
(1-5) {\rm TeV} & ~:~ \tan\beta = 3,\\
(100-500)\GEV & ~:~ \tan\beta = 30,
\end{array} \right.
\label{value-mu_H}
\eeq
and
\beq
m_0 =
 (100-250)\GEV ~:~ \tan \beta =3-30.
\label{value-m0}
\eeq
With such large $\mu_H$, the radiative electroweak symmetry
breaking~\cite{PTP68-927} hardly occurs as long as the universal
soft SUSY-breaking mass in eq. (\ref{universalmass}) is imposed. A
solution to this difficulty is to abandon the GUT-like relation among
the gaugino masses in eq. (\ref{GUT-relation}) so that the gluino
gives rise to much larger masses for squarks than those for sleptons
at the electroweak scale.

\noindent
{\it ii}) One may expect that the hierarchy in the stau masses
($m_{\tilde{\tau}_1}\ll m_{\tilde{\tau}_2}$) will give a large
contribution to $\rho$-parameter.  We find, however, that their
contribution is not substantial as far as the mass of the heavier stau
is less than 500 GeV for $m_{\tilde \tau_1}=50\GEV$.

\noindent
{\it iii}) The large $\mu_H\tan\beta$ induces also a large left-right
mixing in the sbottom sector.  We find that in some of the parameter
space in eqs. (\ref{value-mu_H}) and (\ref{value-m0}), the lighter
sbottom mass $m_{\tilde{b}_1}$ may come in the region excluded
experimentally ($m_{\tilde{b}_1}\lsim 45\GEV$).  However, this problem
can be easily solved by giving the
gluino a mass larger  than  expected from the GUT-like relation
(\ref{GUT-relation}), since in this case the lighter sbottom can be
lifted above $ 45\GEV$ through the radiative corrections.  This is
also favorable for suppressing the $b \rightarrow s \gamma$ decay
substantially as will be discussed in ref.~\cite{future}.

\noindent
 {\it iv}) As pointed out in ref.~\cite{PLB338-212}, the similar
lepton-flavor violation may also occur in the framework of SUSY SU(5)
GUTs.  It should be noted here that in the present
model  the soft SUSY breaking masses $m_{L,ij}^2$ for left-handed
sleptons receive significant radiative corrections from the new Yukawa
couplings $y_{\nu, ij}$, while in the SUSY SU(5) case, those for
right-handed sleptons, $m_{E,ij}^2$, are subject to the large
renormalization effects.  Thus, in the present model, the $i=j=3$
element of $m_{L,ij}^2$ becomes smaller than other diagonal elements,
$m_{L,11}^2$ and $m_{L,22}^2$, by amount of
$O((10-30)\%)$~\cite{PLB321-56} as was shown in eq.(\ref{m_L^2@mz}), whereas in
the SUSY SU(5) GUTs the similar mass shift appears in the
right-handed slepton sector. Therefore, if the
soft SUSY-breaking parameters are precisely determined in future
experiments, these two scenarios may be distinguished.

\hspace*{-\parindent}
{\bf Note added}

After completing this work we became aware that in the very recent
paper~\cite{hep-ph/9501334} the lepton-flavor violation due to the
Yukawa couplings $y_{\nu, ij}$ are also examined in the context of
SUSY SO(10) grand unification.  The $l_i\rightarrow l_j\gamma$ process
is calculated there by using the $m_{LR}^2$ mass-insertion, and our
formula in eq. (\ref{ampLR-mass-insertion}) is consistent with their
result of the $y_{\nu, ij}$ Yukawa coupling effects.  This is, however, only
an approximation of our full formula (\ref{ampLR}) in the small
$m_{LR}^2$ region.

\newpage
%
%
\newcommand{\Journal}[4]{{\sl #1} {\bf #2} {(#3)} {#4}}
\newcommand{\APJ}{Ap. J.}
\newcommand{\CJP}{Can. J. Phys.}
\newcommand{\NC}{Nuovo Cimento}
\newcommand{\NP}{Nucl. Phys.}
\newcommand{\PL}{Phys. Lett.}
\newcommand{\PR}{Phys. Rev.}
\newcommand{\PRep}{Phys. Rep.}
\newcommand{\PRL}{Phys. Rev. Lett.}
\newcommand{\PTP}{Prog. Theor. Phys.}
\newcommand{\SJNP}{Sov. J. Nucl. Phys.}
\newcommand{\ZP}{Z. Phys.}

%
%
\begin{figure}[p]
\epsfxsize=15cm
\caption{Feynman diagrams which give rise to $l_i\rightarrow l_j\gamma$.
In each diagram, the blob indicates the flavor-violating mass
insertion of the left-handed slepton and at the cross mark the
external lepton flips its chirality.   The symbols
$\tilde{e}_{Li}$, $\tilde {\nu}_{Li}$, $\tilde{B}$, $\tilde{W}_{3}$, and
$\tilde{W}^{-}$ represent left-handed charged sleptons, left-handed
sneutrinos, bino, neutral wino, and charged wino, respectively. }
\label{fig:feyn_LL}
\end{figure}
%
%
\begin{figure}[p]
\epsfxsize=15cm
\caption{Feynman diagram which gives rise to $l_i\rightarrow l_j\gamma$.
The blobs indicate insertions of the flavor-violating mass
($m_{L,ij}^2$) and the left-right mixing mass ($m_{LR,ii}^2$), and at
the cross mark chirality flip of the bino ($\tilde B$) occurs.  The
symbols $\tilde{e}_{Li}$ and $\tilde{e}_{Ri}$ represent left-handed
charged sleptons and right-handed charged sleptons, respectively.
Notice, however, that we do not use the mass-insertion method in our
calculation as stressed in the text.  }
\label{fig:feyn_LR}
\end{figure}
%
%
\begin{figure}[p]
\epsfxsize=15cm
\caption{Branching ratios for the processes $\tau\rightarrow\mu\gamma$
and $\mu\rightarrow e\gamma$ as  functions of $m_{\tilde{\tau}_2}$.
The solid lines correspond to $Br(\tau \rightarrow \mu \gamma)$ and
the dashed lines to $Br(\mu \rightarrow e \gamma)$.  Here,
we have taken $m_{G2}=45\GEV$ and $m_{\tilde{\tau}_1}=50\GEV$.  We also show
the present experimental upper bounds for each processes by the solid
lines with hatches.  }
\label{fig:br_light}
\end{figure}
%
%
\begin{figure}[p]
\epsfxsize=15cm
\caption{Same as fig.~\protect\ref{fig:br_light} except for
$m_{G2}=90\GEV$ and $m_{\tilde{\tau}_2}=100\GEV$.}
\label{fig:br_heavy}
\end{figure}

\end{document}